\documentstyle[12pt]{article}
\begin{document}
\title{Composite p-branes  in Various Dimensions}
\author{
I.Ya.Aref'eva \\ {\it Steklov Mathematical Institute, 
Vavilov St.42, 117966, Moscow,} \\
        K.S.Viswanathan \\ {\it Department of Physics,
                      Simon Fraser University,}\\ 
    {\it   Burnaby, British Columbia, V5A 1S6, Canada,}\\
       A.I.Volovich \\ {\it
Department of Physics,$~$ Moscow State
 University, }\\ {\it  
Vorobjevi gori, Moscow, 119899;
}\\ {\it  
  Mathematical College, Independent
  University of Moscow,  }\\ {\it  
121108, po box 68, Moscow, Russia},\\ 
   I.V.Volovich \\ {\it Steklov Mathematical
Institute,  
Vavilov St.42, 117966, Moscow}
}
\date{$~$}
\maketitle

\begin{abstract}
We review  an algebraic method of finding
the composite p-brane solutions for a
generic Lagrangian, in arbitrary spacetime dimension,
describing an interaction of a graviton,
a dilaton and one or two antisymmetric tensors.
We set the Fock--De Donder
harmonic gauge for the metric and the "no-force" condition
for the matter fields. Then equations for the antisymmetric
field are reduced to the Laplace
equation and the equation of motion for the dilaton and the
Einstein equations for the metric are reduced to an algebraic
equation.
 Solutions  composed of $n$ constituent p-branes with
$n$ independent harmonic functions are  given.
The form of the solutions demonstrates the harmonic
functions superposition rule in diverse dimensions.
Relations  with known solutions in $D=10$ and
$D=11$ dimensions are discussed.

\end{abstract}

\newpage
 \section{Introduction}
Recent remarkable developments in superstring theory
led to the discovery that the five known superstring
theories in ten dimensions are related by duality
transformations  and to the conjecture that
$M$-theory
underlying the superstring theories
and $11$ dimensional supergravity exists \cite {FILQ}-\cite{Vaf}.
Duality requires  the presence of extremal black holes
in the superstring spectra. A derivation of the Bekenstein-Hawking
formula for the entropy of certain extreme black holes was given
by using the D-brane approach \cite{SV}-\cite{pol1}.

In all these developments
the study of p-brane solutions of the supergravity equations
play an important role \cite{DGHR}-\cite{berg}.
To clarify the general picture of p-brane solutions
it seems useful to have solutions in arbitrary spacetime dimension.

In this talk results obtained in our recent works
\cite{AV,AVV,IN} will be
presented.  We use a systematic algebraic method of finding
$p$-brane solutions in diverse dimensions.  We start from
an ansatz for  a metric on a product manifold and use
the Fock--De Donder harmonic gauge.  It leads to a  simple form
for the Ricci tensor.  Then we consider an ansatz for the matter
fields and use the "no-force"  condition. This leads to
a simple form of the stress-energy
tensor and moreover the equation for the antisymmetric fields
are reduced to the Laplace equation. The Einstein equations
for the metric and the equation of motion for the dilaton
under these conditions are reduced thence to an algebraic equation
for the parameters in the Lagrangian.

\section{Ricci tensor in the Fock--De Donder gauge}
Let us be given a $D$-dimensional manifold of the product
form $M^D=M^q\times M^{r_1}\times M^{r_2}\times ...
\times M^{r_n}\times M^{s+2}$.
We will use the following ansatz for a metric
on $M^D$:

\begin{equation}
                           \label{2}
ds^{2}=e^{2A}\eta_{\mu \nu} dy^{\mu}
dy^{\nu}+\sum _{i=1}^{n}e^{2F_{i}}(dz_{i}^{m_{i}})^{2}
+e^{2B}(dx^{\gamma})^{2}
\end{equation}
Here $\eta_{\mu\nu}$
is a  Minkowski metric, $\mu, \nu =0,...,q-1;\ \  m_{i}=1,...,r_i;\ \
 \gamma=1,...,s+2$,
\begin{equation}
                           \label{3}
q+\sum _{i=1}^{n}r_{i}+ s+2=D
\end{equation}
and the functions $A, F_i $ and $B$ depend only on $x$.
The metric (\ref{2}) is the sum of $n+2$ blocks.
We shall call it
the $n+2$-block $p$-brane metric.

The Ricci tensor for the above  metric reads
$$R_{\mu\nu}=-\eta _{\mu\nu}e^{2(A-B)}[\Delta  A
+q(\partial A)^{2}+$$
$$
 \sum _{i=1}^{N}r_{i}(\partial A \partial F_{i})]
+s(\partial A \partial B)],
$$
$$R_{m_{i}n_{i}}=-\delta_{m_{i}n_{i}}e^{2(F_{i}-B)}[\Delta  F_{i}
+q(\partial A\partial F_{i})+$$
$$\sum _{j=1}^{N}r_{j}(\partial F_{j} \partial
F_{i})] +s(\partial F_{i} \partial B)],
$$
$$R_{\alpha\beta}=-q\partial_{\alpha} \partial_{\beta}A
-\sum _{i=1}^{N}r_{i}\partial_{\alpha} \partial_{\beta} F_{i}-
s\partial_{\alpha} \partial_{\beta}B -
$$
$$
-q
\partial_{\alpha} A\partial_{\beta} A - \sum _{i} r_{i}\partial_{\alpha}
F_{i}\partial_{\beta} F_{i} +$$
$$
s\partial_{\alpha} B\partial_{\beta} B
+ q(\partial_{\alpha} A\partial_{\beta}
B+ \partial_{\alpha} B\partial_{\beta} A)+$$
$$
\sum _{i} r_{i}(\partial_{\alpha}
B\partial_{\beta} F_{i} +
\partial_{\alpha}
 F_{i}\partial_{\beta}B)-$$
$$
\delta_{\alpha\beta}[\Delta B+q(\partial A \partial B)+
\sum _{i=1}^{N}r_{i}(\partial B\partial F_{i})]
+s(\partial B)^{2}]
$$
We shall use the Fock-De Donder gauge
\begin{equation}
\partial_M(\sqrt{-g}g^{MN})=0
                                        \label{9a}
\end{equation}
For the metric (\ref{2}) this leads to the following
important condition
\begin{equation}
qA+\sum r_{i}F_{i} +sB=0
                                        \label{9}
\end{equation}
Then the Ricci tensor for the above metric  takes the simple  form
 \begin{equation}
 R_{\mu\nu}=-\eta_{\mu\nu}e^{2(A-B)}\Delta  A,
 \label{10}
\end{equation}
 \begin{equation}
 R_{m_{i}n_{i}}=-\delta_{m_{i}n_{i}}e^{2(F_{i}-B)}\Delta  F_{i}
 \label{10a}
\end{equation}
\begin{eqnarray}
 R_{\alpha\beta}=- q\partial_{\alpha} A\partial_{\beta} A -
\sum _{i} r_{i}\partial_{\alpha} F_{i}\partial_{\beta} F_{i}
+                                                         \nonumber\\
s\partial_{\alpha} B\partial_{\beta} B
-\delta_{\alpha\beta}\Delta B~~~~~~~~~~~~~~
                              \label{12}
\end{eqnarray}
We shall consider the Einstein equations in the form
\begin{equation}
                                         \label{EE}
R_{MN}={\cal G}_{MN}
\end{equation}
\begin{equation}
                                         \label{EE1}
{\cal G}_{MN}=T_{MN}-\frac {1}{D-2}g_{MN}T
\end{equation}

\section{$n+2$-block solution}

In this section we shall consider the following action
\begin{equation}
                           \label{1}
 I=\int d^{D}x\sqrt{-g}
 (R-\frac{(\nabla \phi)^2}{2} -\frac{e^{-\alpha \phi}}{2(d+1)!}F^{2}_{d+1})
\end{equation}
It describes the interaction of the gravitation field $g_{MN}$
with the dilaton $\phi$ and with one antisymmetric field:
$F_{d+1} $ is a closed $d+1$-differential form.
The stress energy tensor for the action (\ref{1}) is
$$
 T_{MN}= \frac{1}{2}
  (\partial _{M}\phi \partial _{N}\phi -
  \frac{1}{2}g_{MN} (\partial \phi)^{2}) +
$$
\begin{equation}
   \frac{e^{-\alpha\phi}}{2d!}
   (F_{MM_{1}...M_{d}}F_{N}^{M_{1}...M_{d}}-
 \frac{g_{MN}}{2(d+1)}F^{2})
                                          \label{15}
\end{equation}

In this section we shall consider an electric ansatz for the action
(\ref{1}) which leads to a $n+2$-block metric.
Let us take the ansatz (\ref{2}) with $r_{i}=r>1$.
Then (\ref{3}) takes the form
\begin{equation}
D=q+nr+s+2
                                          \label{15s}
\end{equation}
We use
the  following
ansatz for the $d$-form  ${\cal A}$
\begin{equation}
{\cal A} =\omega _{0}^q\wedge[
\omega _{1}^{r}h_{1}e^{C_{1}}+
...+
\omega _{n}^{r}h_{n}e^{C_{n}}]
                                                       \label{21}
\end{equation}
where  $d=q+r$,
$$\omega _{0}=dy^{0}\wedge dy^{1}\wedge ...\wedge dy^{q-1}$$
$$
\omega _{i}^{r}=dz_{i}^{1}\wedge ...\wedge dz_{i}^{r},$$
$C_{i}$ are functions of $x$ and $h_{i}$ are some constants.
More general ansatz has been considered in
\cite{AR,IM}.
For the stress-energy tensor we get
$$
T_{\mu \nu}=-\eta_{\mu \nu} e^{2(A-B)}[\frac{1}{4}(\partial \phi)^{2}+
\sum _{i=1}^{n}
\frac{h_{i}^{2}}{4}e^{-2qA-2rF_{i}
-2C_{i}}(\partial C_{i})^{2}],
$$
$$
T_{m_{i}n_{i}}=\delta_{m_{i}n_{i}} e^{2(F_{i}-B)}
[\frac{1}{4}(\partial \phi)^{2}-$$
$$
\frac{h_{i}^{2}}{4}e^{-2qA-2rF_{i}+2C_{i}}(\partial C_{i})^{2}+
\sum _{j\neq i}^{n}\frac{h_{j}^{2}}{4}e^{-2qA-2rF_{j}+2C_{j}}
(\partial C_{j})^{2}],
$$
$$T_{\alpha \beta}=\frac{1}{2}[\partial _{\alpha}
\partial _{\beta}\phi-\frac{1}{2}
\delta _{\alpha\beta}(\partial \phi)^{2}]-$$
$$
\sum _{i=1}^{n}\frac{h_{i}^{2}}{2}e^{-2qA-2rF_{i}+2C_{i}}
[\partial_{\alpha} C_{i}\partial_{\beta} C_{i}
-\frac{\delta_{\alpha \beta} }{2}
(\partial C_{i})^{2}]
$$
We set the following ("no-force") condition
\begin{equation}
-\alpha \phi -2qA-2rF_{i}-2C_{i}=0,~~i=1,...n
                                              \label{27}
\end{equation}
Then the form of $T_{MN}$   simplifies  drastically
and the Einstein equations (\ref{EE}) take the form
\begin{equation}
 \Delta  A =\sum _{i}t
 h^{2}_{i}(\partial C_{i})^{2},
                                               \label{2.33}
\end{equation}
\begin{equation}
\Delta  F _{i}=t h^{2}_{i}
 (\partial C_{i})^{2}
-\sum _{j\neq i}u h^{2}_{j}(\partial C_{j})^{2},
                                           \label{2.34}
\end{equation}
$$
- q\partial_{\alpha} A\partial_{\beta} A -
\sum _{i} r_{i}\partial_{\alpha} F_{i}\partial_{\beta} F_{i}
~~~~~~~~~~$$
$$
-s\partial_{\alpha} B\partial_{\beta} B
- \delta_{\alpha\beta}\Delta B=
\frac{1}{2}\partial _{\alpha}\phi
\partial _{\beta} \phi -
~~~~~~~~~~~$$
\begin{equation}
\sum _{i=1}^{n}h^{2}_{i}[\frac{1}{2}
\partial _{\alpha} C_{i}\partial _{\beta} C_{i}-u\delta_{\alpha \beta}
(\partial C_{i})^{2}],
                                         \label{2.35}
\end{equation}
where
\begin{equation}
                                          \label{2.31}
t =\frac{D-2-q -r }
{2(D -2)},~~
u =\frac{q +r }
{2(D -2)}.
\end{equation}
The equation of motion for the antisymmetric field,
\begin{equation}
\partial _{M}(\sqrt{-g}F^{MM_{1}...M_{d}})=0,
                                           \label{16}
\end{equation}
 under conditions
(\ref{9}) and (\ref{27})
for the ansatz (\ref{21})
reduces to the Laplace equation
\begin{equation}
\partial _{\alpha}\partial _{\alpha}(e^{ -C_{i}} )=0~
\mbox{or}~ \Delta C_i=(\partial C_i)^{2}.
                                         \label{36}
\end{equation}
Equation of motion for the dilaton reads
\begin{equation}
\Delta \phi -\frac{\alpha}{2}\sum _{i}
 h^{2}_{i}(\partial C_{i})^{2} = 0
                                         \label{2.36}
\end{equation}
or by using (\ref{36})
\begin{equation}
\Delta (\phi -\frac{\alpha}{2}\sum _{i}
 h^{2}_{i}C_{i}) = 0
                                         \label{2.36a}
\end{equation}

We take the following solution of (\ref{2.36a})
\begin{equation}
\phi=\frac{\alpha}{2}\sum _{i}h^{2}_{i} C_{i}
                                               \label{2.37}
\end{equation}
Analogously we take the following solutions of
 (\ref{2.33}) and  (\ref{2.34})
\begin{equation}
 A=t\sum _{i}h_{i}^{2} C_{i},
                                               \label{37}
\end{equation}
\begin{equation}
F_{i}=th_{i}^{2} C_{i}
-u\sum _{j\neq i}h_{j}^{2} C_{j}
                                      \label{38}
\end{equation}
To cancel out the terms proportional to
$\delta_{\alpha \beta}$ in (\ref{2.35}) we set
\begin{equation}
B =-u \sum _{i} h^{2}_{i} C_{i}
                                         \label{2.39}
\end{equation}
Note that  if we want  to solve equations  for arbitrary
harmonic functions $H_{i}=$ $\exp (-C_{i})$ we have also to assume the
conditions which follow  from (\ref{27}), (\ref{9}) and nondiagonal
part of (\ref{2.35}). Equations  (\ref{27}) lead to the relations
$$[\frac{\alpha ^{2}}{4}
+(q+r)t ]h_{i}^{2} C_{i}+
$$
\begin{equation}
                                \label{2.40.0}
[\frac{\alpha ^{2}}{4}+qt-
r u ]\sum _{j\neq i} h_{j}^{2} C_{j}
=C_{i},~~~i=1,...n.
\end{equation}
If $n\neq 1$ then under the assumption of independence of $C_{i}$
the relations (\ref{2.40.0}) yield
\begin{equation}
\frac{\alpha ^{2}}{4}=r u -
q t ,
                                      \label{2.40}
\end{equation}
                            \begin{equation}
[\frac{\alpha^{2}}{4}+(q +r )t ]
h^{2}_{i} =1.
                                      \label{2.41}
\end{equation}
If $n=1$ we get only (\ref{2.41}).
Since $t $ and $u $
are given by (\ref{2.31})
the condition (\ref{2.40}) leads to the following equation
\begin{equation}
(\frac{\alpha ^{2}}{2}+q )(D -2)
=d^{2}
                                         \label{RR}
\end{equation}
Equation (\ref{RR}) plays the central role in our approach.
For given parameters $D,d$ and $\alpha$ in the Lagrangian (\ref{1})
we have to find a positive integer $q$ which solves equation (\ref{RR}).
In this sense we can interpret equation (\ref{RR}) as "quantization"
of the parameter $\alpha$ in the Lagrangian (see also \cite{AV,Berg}).

Note that under this condition the formula (\ref{2.31}) takes the form
\begin{equation}
u =\frac{2q +\alpha ^{2}}
{4(q +r) },~~
t =\frac{2r -\alpha ^{2}}
{4(q +r) }
                                         \label{2.31'}
\end{equation}

The LHS of (\ref{2.41}) for $t $  and $u $
given by formulae (\ref{2.31'})
can be represented as $r h^{2}/2$, and , therefore,
equation (\ref{2.41}) gives
\begin{equation}
h^{2}_{i} =h^{2}\equiv \frac{2}{r},
                                      \label{2.43}
\end{equation}

By straitforward calculations one can check that for
$\alpha$, $q $, $r $ and $D $
satisfying relation (\ref{RR}) and $h $ given by
(\ref{2.43}), equations (\ref{9}) as well as equations
giving a compensation of terms
$\partial_{\alpha} C_{j}\partial_{\beta} C_{i}$
in the both sides of equation (\ref{2.35}) are fulfilled.

This calculation shows that the  metric
\begin{equation}
ds^{2}=(H_{1}H_{2}...H_{n})^{\frac{4u}{r }}
\label{2.50.0}
\end{equation}
$$
[
(H_{1}H_{2}...H_{n})^{-\frac{2}{r }}\eta_{\mu \nu} dy^{\mu}dy^{\nu}+
\sum _{i=1}^{n}H_{i}^{-\frac{2}{r }}(dz^{m_{i}})^{2}
+(dx^{\gamma})^{2}],$$
 and  matter fields in the form
\begin{equation}
\exp \phi=(H_{1}H_{2}...H_{n})^{-\alpha /r }
 \label{2.51}
\end{equation}
\begin{equation}
{\cal A} =\sqrt{\frac{2}{r}}
\omega _{0}\wedge[
\omega _{1}H^{-1}_{1}+...
+\omega _{n} H^{-1}_{n}]
                                                       \label{210}
\end{equation}
is a solution of the theory (\ref{1})
if $H_{i}(x)$, $i=1,...n$, are harmonic functions and
the parameters in the Lagrangian $D$, $d$ and  $\alpha $
are such that equations (\ref{15s}) and (\ref{RR}), i.e.
$$d=q+r,~~D=q+nr+s+2,~~~~~~~~~~~~~~~~~~~~~$$
\begin{equation}
(\frac{\alpha ^{2}}{2}+q )(D -2)
=d^{2}
                                         \label{15ss}
\end{equation}
admit
solutions with positive integers $q,n,r$ and $s+2$.

Let us note that the formula (\ref{2.50.0}) proves the 
harmonic superpositon rule for the ansatz (\ref{21}).
Indeed, $u$ and $r$ can be written as

$$u=\frac{d}{2(D-2)},~~~r=d(1-\frac{d}{D-2})+\frac{{\alpha}^{2}}{2}$$
One can easily see that the exponents in the formula are
defined by the two-block solution (\cite{Lu})
$$ds^2=H^{\frac{4u}{r}}[H^{-\frac{2}{r}}\eta_{\hat{\mu}\hat{\nu}}
dy^{\hat{\mu}}dy^{\hat{\nu}}
+dx^{\hat{\gamma}}dx^{\hat{\gamma}}]$$
$\hat{\mu},\hat{\nu}=1,...d-1,~~~\hat{\gamma}=1,...D-d$.
Therefore, having the simplest two-block solution one can produce
n-block solution. Let us note that this prescription works only
for the case when the characteristic equation (\ref{RR}) is
satisfied. The harmonic function rule was formulated in (\cite{Ts212}).
for the case $D=10,~D=11$. The formula (\ref{2.50.0})
demonstrates that the harmonic superposition rule
holds for arbitrary dimension.

\subsection{Examples}
\subsubsection {$\alpha =0$}
Note that equations (\ref{15ss}) are very restrictive
since it has
to be solved for integers $q,d,r$ and $D$.
Let us present some  examples.

For dimensions $D=4,5,6,7,8$  and $9$  there are solutions only
with $r=0$
and we have 2-block solutions with   $s=0$. In these cases, either the
spacetime is asymptotically $M^{q}\times Y$, where $Y$ is a two-dimensional
conical space, or the metric exhibits logarithmic behaviour
as $|x|\to \infty$ \cite{berg}.

We get more interesting structures in $D=6$ case.
There are four types of solutions of
(\ref{15ss}) with $\alpha =0$:
$$i)~q=1,~ r=1, n=2, s=1,$$
$$ii)~q=1,~ r=1, n=3, s=0,$$
$$iii)~q=1,~ r=1, n=4, s=-1,$$
$$ iv)~~q=1,~ r=1, n=5, s=-2.$$
Here we have to assume that different branches with $r=1$
correspond to different
gauge field ${\cal A}^{(I)}=h dy^{0}$$
\wedge dz_{i}H_{i}^{-1}\delta _{iI}$ , $I=1,...n$
(othewise we cannot guarantee the diagonal form of the
stress-energy
tensor). The solution iv) is identified with the Minkowski
vacuum of the  theory.
The solution iii) separates the 6-dimensional
space-time into two asymptotic regions like a domain wall \cite{Lu}.
 The metric for the solution with $s=0$ has a logarithmic behaviour,
$H=\sum _{a}
\ln (Q_{a}/|x-x_{a}|^{2})$.

For $s=1$ we have
$$ds^{2}=(H_{1}H_{2})[-(H_{1}H_{2})^{-2}dy_{0}^{2}+$$
\begin{equation}
 \label{D6}
H_{1}^{-2}dz_{1}^{2}
                    +H_{2}^{-2}dz_{2}^{2}+(dx_{i})^{2}] ,
~~       i=1,2,3
\end{equation}
\begin{equation}
 \label{D6f}
{\cal A}^{(1)}=\sqrt{2}dy_{0}\wedge dz_{1} H_{1}^{-1}, ~
~{\cal A}^{(2)}=\sqrt{2}dy_{0}\wedge dz_{2} H_{2}^{-1}
\end{equation}
\begin{equation}
 \label{Har1}
H_{1}=1+\sum_{a=1}^{l_{1}} \frac{Q^{(1)}_{a}}{|x-x^{(1)}_{a}|};
~~  H_{2}=1+\sum_{b=1}^{l_{2}} \frac{Q^{(2)}_{b}}{|x-x^{(2)}_{b}|},
 \end{equation}
$$|x-x_{a}|=(\sum _{i=1}^{3}|x_{_{i}}-x_{a_{i}}|^{2})^{1/2}.
~~~~~~~~~~~~~~~~~~~~~~~~~~~~~~~~~$$
If $l_{1}=l_{2}=l$ and $x^{(1)}_{a}=x^{(2)}_{a}=x_{a}$
the metric (\ref{D6}) has horizons at the points
$x_{a}$.  The area
(per unit of length in all p-brane directions)
of the horizons
$x=x_{a}$  is
\begin{equation}
 \label{H6}
A_{4}= 4\pi \sum _{a=1}^{l}Q_{a}^{(1)} Q_{a}^{(2)}
\end{equation}
This confirms an observation \cite{KLOPP,kl} that extremal black
holes have non vanishing event horizon in the presence of two or more charges
(electric or magnetic).
There are more solutions for several scalar fields \cite{Em,kl,Lust}.

For $D=10 $ we have two solutions with $s=0$.

$~i)~~~ q=8,~ r=0, s=0, n=1$;

$~ii)~ q=2,~ r=2,~ s=0,~n=3.$

There is also solution with
$q=2,~ r=2,~ s=2,~n=2,$
$$
ds^{2}=(H_{1}H_{2})^{1/2}[(H_{1}H_{2})^{-1}(-dy_{0}^{2}+dy_{1}^{2}+
Kdu^{2})
$$
\begin{equation}
 \label{D10}
+H_{1}^{-1}(dz_{1}^{2}+dz_{2}^{2})+H_{2}^{-1}(dz_{3}^{2}+dz_{4}^{2})
+(dx_{i})^{2}] ,
 \end{equation}
    $$   H_{1}=1+\sum_{a} \frac{Q^{(1)}_{a}}{|x-x_{a}|^{2}}; ~
    H_{2}=1+\sum_{b} \frac{Q^{(2)}_{b}}{|x-x_{b}|^{2}},$$
    $$   K=\sum_{a} \frac{Q_{a}}{|x-x_{a}|^{2}};~~u=y_{0}+y$$
The area of this horizon
is
\begin{equation}
 \label{H10}
A_{8}= \omega _{3}\sum _{a} (Q^{(1)}_{a} Q^{(2)}_{a}Q_{a})^{1/2}
\end{equation}
where $\omega_{3}=2\pi ^{2} $ is   the area  of the  unit 3-dimension sphere.
In this case the different components of the same
gauge field act as fields corresponding to different charges.

For $D=11$  we have the following solution with $s>0$.\\
$i)~~q=1,~ r=2, s=2,~n=3,$
$$
ds^{2}=(H_{1}H_{2}H_{3})^{1/3}[-(H_{1}H_{2}H_{3})^{-1}dy_{0}^{2}
+H_{1}^{-1}(dz_{1}^{2}+$$
\begin{equation}
 \label{D10'}
dz_{2}^{2})+H_{2}^{-1}(dz_{3}^{2}+dz_{4}^{2})
+H_{3}^{-1}(dz_{5}^{2}+dz_{6}^{2})+dx_{i}^{2}],
 \end{equation}
  $$   H_{c}=1+\sum_{a} \frac{Q^{(c)}_{a}}{|x-x_{a}|^{2}};
    ~~c=1,2,3. $$
For $H_{3}=1$ this solution reproduces  a solution found in \cite{Ts212}.
We get non-zero area of the horizons
$x=x_{a}$
\begin{equation}
 \label{H11}
A_{9}= \omega _{3}\sum _{a} (Q^{(1)}_{a} Q^{(2)}_{a}Q^{(3)}_{a})^{1/2}
\end{equation}

$~ii) ~q=4,~ r=2, s=1,~n=2,$
$$
ds^{2}=(H_{1}H_{2})^{2/3}[(H_{1}H_{2})^{-1}(-dy_{0}^{2}+dy_{1}^{2}
+dy_{2}^{2}+
dy_{3}^{2})+$$
\begin{equation}
 \label{D11}
H_{1}^{-1}(dz_{1}^{2}+dz_{2}^{2})+H_{2}^{-1}(dz_{3}^{2}+dz_{4}^{2})
+\sum _{i=1}^{3}dx_{i}^{2}],
\end{equation}
where   $  H_{1}$ and $H_{2}$ are given by
(\ref{Har1}). This solution has been recently  found in
\cite{Ts212}.
The area of the horizon $x_{a}=x_{b}$ is equal to zero.

\subsubsection {$\alpha \neq 0$}

Let us present some  examples of solution of equation (\ref{RR}).
For $D=10$ we have  $q=1,~ r=3,$ $\alpha =\pm \sqrt{2}, s=1,$
$n=2$
and the corresponding metric has the form
$$
ds^{2}=(H_{1}H_{2})^{1/3}[-H_{1}H_{2})^{-2/3}dy_{0}^{2}+
H_{1}^{2/3}(dz_{1}^{2}
+dz_{2}^{2}+dz_{3}^{2})+$$
\begin{equation}
 \label{D11'}
H_{2}^{-2/3}(dz_{4}^{2}+dz_{5}^{2}+dz_{6}^{2})
+\sum_{i=1}^{3}dx_{i}^{2}],
\end{equation}
where   $  H_{1}$ and $H_{2}$ are given by
(\ref{Har1}). This solution corresponds to IIA supergravity.
The area of the horizon $x=x_{a}=x_{b}$ is equal to zero.

\section{Three block solution}

Let us consider the following action
with two gauge fields
\begin{eqnarray}
 I=\int d^{D}x\sqrt{-g}
 (R-\frac{1}{2}(\nabla  \phi) ^{2}
 -                                                        \nonumber\\
\frac{1}{2(q+1)!}e^{-\alpha \phi}F^{2}_{q+1}
-\frac{1}{2(d+1)!}e^{\beta \phi}G^{2}_{d+1}).
      \label{011}
\end{eqnarray}
Here $F_{q+1} $ is a closed $q+1$-differential form and  $G_{d+1}$
is a closed $d+1$-differential form.

There is the following solution \cite{AVV}
$$ds^{2}=H_{1}^{-2a_{1}}H_{2}^{-2a_{2}}\eta_{\mu \nu} dy^{\mu} dy^{\nu}+
$$
\begin{equation}
H_{1}^{-2b_{1}}H_{2}^{-2a_{2}}
dz_{n}^{2}
+H_{1}^{-2b_{1}}H_{2}^{-2b_{2}}dx_{\gamma}^{2},
 \label{s2f}
\end{equation}
where $\eta_{\mu\nu}$ is a flat Minkowski metric,
$\mu, ~\nu = 0,...,q-1;$ $m,n=1,2,...,d-q,$ and
$\gamma=1,...,D-d.$
For definitness we assume  that $ D>d\geq q$.

The parameters $a_{i}$ and
 $b_{i}$ in the solution (\ref{s2f}) are rational functions of
the parameters in the action (\ref{011}):
\begin{equation}
a_{1}=\frac{2\tilde q}{\alpha^{2}(D-2)+2q\tilde q},
~~b_{1}=-\frac{q}{\tilde{q}}a_{1},
 \label{13}
\end{equation}
\begin{equation}
a_{2}=\frac{\alpha^{2}(D-2)}
{\alpha^{2}d(D-2)+2\tilde d q^{2}},
~~b_{2}=-\frac{d}{\tilde {d}}a_{2},
\label{14}
\end{equation}
where
\begin{equation}
\tilde d=D-d-2,~~~~\tilde q=D-q-2.
\label{14a}
\end{equation}

The solution  (\ref{s2f}) is valid only if the following
relation between parameters in the action is satisfied
\begin{equation}
\alpha\beta=\frac{2q\tilde d}{D-2}
\label{RRR}
\end{equation}
There are   two arbitrary harmonic functions
  $H_{1}$ and $H_{2}$ of variables $x^{\gamma}$ in (\ref{s2f}),
 \begin{equation}
  \Delta H_{1}=0,~~~~~\Delta H_{2}=0.
  \label{HF}
 \end{equation}

Non-vanishing components of the differential form are given by
 \begin{equation}
  {\cal A}_{\mu_{1}...\mu_{q}}=h
  {\epsilon_{\mu_{1}...\mu_{q}}}H_{1}^{-1},~~~~ F=d{\cal A},
  \label{18}
 \end{equation}
 \begin{equation}
  {\cal B}_{I_{1}...I_{d}}=k
  \epsilon_{I_{1}...I_{d}}H_{2}^{-1}, ~~~G=d{\cal B},.
  \label{18a}
 \end{equation}
Here $I=0,...d-1$, $\epsilon _{123..,q}=1$, $\epsilon _{123...d}=1$ and
$h$ and $k$ are given by the formulae
\begin{equation}
      h^{2}=\frac{4(D-2)}{\alpha ^{2}(D-2)+2q{\tilde q}},
                         \label{194}
\end{equation}
\begin{equation}
k^{2}=\frac{2\alpha^{2} (D-2)^{2}}
{\tilde{d}[\alpha ^{2}d(D-2)+2q^{2}\tilde{d}]},
 \label{195n}
\end{equation}

The dilaton field is
\begin{equation}
  \phi=\frac{1}{2}\beta k^{2}\ln H_{2}-\frac{1}{2}
  \alpha h^{2}\ln  H_{1}.
  \label{17}
\end{equation}
We have obtained \cite{AVV} the solution  (\ref{s2f}) by reducing
the Einstein
equations to the system of algebraic equations. To this end
we have assumed the "no-force" condition, that is an analog of
relations (\ref{27}).
This condition gives a linear dependence between functions
in the Ansatz (\ref{18}), (\ref{18a}). To satisfy a nonlinear relation
that follows from the Einstein equations we have to assume the
relation (\ref{RRR}).

\subsection{Dual Action}
Let us  consider  the following  "dual" action
$$\tilde I=\int d^{D}x\sqrt{-g}
 (R-\frac{1}{2}(\nabla  \phi) ^{2}
 -$$
\begin{equation}
\frac{e^{-\alpha \phi}}{2(q+1)!}F^{2}_{q+1}
 -\frac{e^{\tilde\beta \phi}}{2(s+1)!}G^{2}_{s+1}),
      \label{11}
 \end{equation}
where  $G_{s+1}$ is a closed $s+1$-differential form.
If $s$ is related to $d$ by
\begin{equation}
s=D-d-2, ~~\mbox{i.e.}~~s=\tilde d
      \label{dr}
 \end{equation}
  and
\begin{equation}
\tilde\beta =-\beta
      \label{11a}
 \end{equation}
then the  solution for the metric (\ref{s2f}) with the differential
form $F$ (\ref{18}) and the dilaton (\ref{17}) is
valid also  for the action (\ref{11}). An expression for
the antisymmetric
field $G$ will be different, namely
 \begin{equation}
   G^{\alpha_{1}...\alpha _{\tilde d+1}}=k
   H_{1}^{\sigma_{1}}H_{2}^{\sigma_{2}}
   \epsilon ^{\alpha _{1}...\alpha_{\tilde d+1}\beta}
   \partial _{\beta} H_{2}^{-1}.
  \label{1999}
 \end{equation}
here $\epsilon ^{123..\tilde d+2}=1$ and
\begin{equation}
\sigma_{1}=\frac{\alpha\beta h^{2}}{2}(1-\frac{1}{s }),
~~~~\sigma_{2}=\frac{\beta k^{2}}{2}(\frac{1}{s }-1)
\label{917}
\end{equation}

Equations of motion for the case of one form corresponds to equation of
motion for ansatz (\ref{12}), (\ref{18}) and (\ref{1999}) for the
dual action (\ref{11}) when $\alpha$=
$\beta$ and $q=s$.
    This three-block p-branes solution
for the Lagrangian with one differential form for various dimensions
of the space-time  was found in \cite{AV}. It
contains previously known D=10 case \cite{Ts212,CM}.

Note that the metric (\ref{s2f}) describes also the solution for the
action with
the form $F_{q+1}$ replaced by its dual $F_{\tilde{q}+1}$ with
$\tilde{q}+q+2=D$  and $\alpha \to \tilde{\alpha}=-\alpha$.
One can also change two forms $F$ and $G$ to their dual version
without changing the metric (\ref{s2f}).

      Below  we consider equations of motion for the action (\ref{011}).
The energy-momentum tensor is for the theory
with the action  (\ref{011}) has the form
$$
 T_{MN}= \frac{1}{2}
  (\partial _{M}\phi \partial _{N}\phi -
  \frac{1}{2}g_{MN} (\partial \phi)^{2})
$$
$$
 + \frac{e^{-\alpha \phi}}{2q!}(F_{MM_{1}...M_{q}}F_{N}^{M_{1}...M_{q}}-
 \frac{g_{MN}}{2(q+1)}F^{2})
$$
$$
 + \frac{e^{-\alpha \phi}}{2s!}(G_{MM_{1}...M_{s}}
  G_{N}^{M_{1}...M_{s}}-
 \frac{g_{MN}}{2(s+2)}G^{2})
$$
The equation of motion for the antisymmetric fields are
\begin{equation}
\partial _{M}(\sqrt{-g}e^{-\alpha \phi}F^{MM_{1}...M_{q}})=0,
                                  \label{111a}
\end{equation}
\begin{equation}
\partial _{M}(\sqrt{-g}e^{-\beta \phi}G^{MM_{1}...M_{s}})=0,
                                  \label{111}
\end{equation}
and one has the Bianchi identity
\begin{equation}
 \epsilon ^{M_{1}...M_{q+2}}\partial_{M_{1}}F_{M_{2}...M_{q+2}}=0.
 \label{112}
\end{equation}
\begin{equation}
 \epsilon ^{M_{1}...M_{s+2}}
 \partial_{M_{1}}G_{M_{2}...M_{s+2}}=0.
 \label{1129}
\end{equation}
The equation of motion for the dilaton is
$$
\partial _{M}(\sqrt{-g}g^{MN}\partial _{N }\phi )
+
\frac{\alpha}{2(q+1)!}\sqrt{-g}e^{-\alpha \phi}F^{2}
+ $$
\begin{equation}
\frac{\beta }{2(s+1)!}\sqrt{-g}e^{-\beta \phi}G^{2}
=0.
 \label{113}
\end{equation}
We shall solve equations (\ref{EE}), (\ref{111a})-(\ref{113})  by using
the following Ansatz for the metric
\begin{equation}
ds^{2}=e^{2A(x)}\eta_{\mu \nu} dy^{\mu} dy^{\nu}+
e^{2F(x)}(dz^{n})^{2}
+e^{2B(x)}(dx^{\gamma})^{2},
 \label{114}
\end{equation}
where $\mu$, $\nu$ = 0,...,q-1, $\eta_{\mu\nu}$
is a flat Minkowski metric, $n$=$1,2,...,r$ and
 $\gamma$ =$1,...,s+2$. Here $A$, $B$
 and $C$ are functions on $x$.
Non-vanishing components of the  differential forms are
\begin{equation}
{\cal A}_{\mu_{1}...\mu_{q}}=h{\epsilon_{\mu_{1}...\mu_{q}}}e^{C(x)},
~F=d{\cal A}
 \label{115}
\end{equation}
\begin{equation}
 G^{\alpha_{1}...\alpha _{s+1}}=\frac{1}{\sqrt{-g}}
 k e^{\beta
 \phi}\epsilon ^{\alpha_{1}...\alpha_{s+1}\gamma}
 \partial _{\gamma} e^{\chi} ,
 \label{116}
\end{equation}
where $h$ and $k$ are constants.
$(\mu \nu)$-components of the energy-momentum tensor for this ansatz
have the form
$$
T_{\mu \nu}=\eta _{\mu \nu}e^{2(A-B)}[-\frac{1}{4}(\partial  \phi )^{2}$$
$$
 -\frac{{h}^{2}}{4}(\partial C)^{2}e^{-\alpha\phi-2qA+2C}
  -\frac{k^{2}}{4}(\partial \chi)^{2}
 e^{2sB+\beta\phi+2\chi} ], $$
$(nm)$-components are:
$$
T_{nm}=\delta _{mn} e^{2(F-B)}[-\frac{1}{4}(\partial
\phi)^{2}+$$
$$\frac{h^{2}}{4}(\partial
 C)^{2}e^{-\alpha\phi-2qA+2C}-
 \frac{k^{2}}{4}(\partial \chi)^{2}e^{2sB+\beta\phi+2\chi}],
$$
and $(\alpha\beta)$-components :
$$T_{\alpha \beta}= \frac{1}{2}[\partial_{\alpha} \phi\partial_{\beta} \phi -
\frac{1}{2}\delta_{\alpha\beta}(\partial\phi)^{2}] - $$
$$
\frac{h^{2}}{2}e^{-\alpha\phi-2qA+2C}[\partial_{\alpha} C\partial_{\beta} C -
\frac{\delta_{\alpha\beta}}{2}(\partial C)^{2}]$$
$$
-\frac{k^{2}}{2} e^{2sB+\beta\phi+
2\chi}[\partial_{\alpha} \chi\partial_{\beta} \chi
 -\frac{\delta_{\alpha\beta}}{2}(\partial \chi)^{2}],
$$
where we use notations
$( \partial A\partial B)=\partial_{\alpha} A\partial_{\alpha} B$
and  $D=q+r+s+2=d+s+2$.

The equations of motion (\ref{111}) for a part of components of the
 antisymmetric field are identically satisfied and for the other part
  they are reduced to a simple equation:
\begin{equation}
\partial _{\alpha}(e^{-\alpha \phi -2qA +C}\partial _{\alpha}C)=0.
                  \label{120}
\end{equation}
For $\alpha$-components of the antisymmetric field we
 also have the Bianchi identity:
\begin{equation}
\partial_{\alpha}(e^{\alpha\phi + 2Bq + \chi }\partial_{\alpha}\chi)=0
  \label{121}
\end{equation}
The equation of motion for the dilaton has the form
$$\partial _{\alpha}(e^{qA +s B + Fr}\partial _{\alpha}\phi )
+\frac{\beta k ^{2}}{2}e^{\beta \phi +2sB+2\chi}
(\partial _{\alpha}\chi)^{2} -
$$
\begin{equation}
\frac{\alpha h^{2}}{2}e^{-\alpha \phi -qA + qB +rF
+2C}(\partial _{\alpha} C)^{2}=0.
  \label{122}
\end{equation}
In order to get rid of  exponents in the above expressions
for the energy-momentum tensor
we impose the following relations:
\begin{equation}
2\chi +2sB +\beta \phi  =0,
                    \label{124}
\end{equation}
\begin{equation}
 2C - 2qA -\alpha\phi = 0.
                       \label{125}
\end{equation}
and we also have
\begin{equation}
 qA+rF+sB  = 0.
                       \label{125a}
\end{equation}
Then the tensor ${\cal G}_{MN}$ (\ref{EE1}) will have the form
$$
{\cal G}_{\mu\nu}=\eta _{\mu\nu}e^{2(A-B)}
[-uh^2(\partial C)^2-vk^2(\partial\chi)^2],$$
$${\cal G}_{mn}=\delta _{mn}e^{2(F-B)}
[th^2(\partial C)^2-vk^2(\partial\chi)^2]$$
$$
{\cal G}_{\alpha\beta}=\frac{1}{2}\partial_{\alpha} \phi\partial_{\beta} \phi
-\frac{h^{2}}{2}\partial_{\alpha} C\partial_{\beta} C
-\frac{k^{2}}{2} \partial_{\alpha} \chi\partial_{\beta} \chi
+$$
$$\delta_{\alpha\beta}[th^2(\partial C)^{2}+wk^2(\partial \chi)^{2}]
$$
Here
\begin{equation}
   u=\frac{1}{2}(1-\frac{q}{D-2}),~~v=\frac{1}{2}-\frac{q+r}{2(D-2)}
 \label{u}
\end{equation}
\begin{equation}
   t=\frac{q}{2(D-2)},~~w=\frac{q+r}{2(D-2)}
 \label{t}
\end{equation}
The Einstein equations (\ref{EE}) under the conditions (\ref{124}),
(\ref{125}) and (\ref{125a}) are crucially simplified and take the form
\begin{equation}
  \label{Emu}
\Delta A=uh^{2}(\partial c)^{2}+vk^{2}(\partial \chi)^{2}
\end{equation}
\begin{equation}
  \label{En}
\Delta F=-th^{2}(\partial c)^{2}+vk^{2}(\partial \chi)^{2}
\end{equation}
$$
- q\partial_{\alpha} A\partial_{\beta} A -
\sum _{i} r_{i}\partial_{\alpha} F_{i}\partial_{\beta} F_{i}
+
s\partial_{\alpha} B\partial_{\beta} B
-$$
$$
\delta_{\alpha\beta}\Delta B=
 -
\frac{h^{2}}{2}[\partial_{\alpha} C\partial_{\beta} C
 -
2t\delta_{\alpha\beta}(\partial C)^{2}]
$$
\begin{equation}
  \label{Eal}
-\frac{k^{2}}{2} [\partial_{\alpha} \chi\partial_{\beta} \chi
 -2w\delta_{\alpha\beta}(\partial \chi)^{2}]
+ \frac{1}{2}\partial_{\alpha} \phi\partial_{\beta} \phi
,
\end{equation}

Now equations (\ref{120}),(\ref{121}) and (\ref{122})
will have the following forms, respectively,
\begin{equation}
\partial _{\alpha}(e^{-C}\partial _{\alpha}C)=0,~~~~~
\partial _{\alpha}(e^{-\chi}\partial _{\alpha}\chi)=0,
                                   \label{126}
\end{equation}
\begin{equation}
\Delta \phi +\frac{\beta k ^{2}}{2}(\partial _{\alpha}  \chi  )^{2}-
\frac{\alpha  h^{2}}{2}(\partial _{\alpha} C )^{2}=0.
                  \label{127}
\end{equation}
One rewrites (\ref{126}) as
\begin{equation}
 \Delta C =(\partial C)^{2},~~~~~
 \Delta \chi =(\partial \chi)^{2}.                  \label{128}
\end{equation}
Therefore (\ref{127}) will have the form
\begin{equation}
\Delta \phi +\frac{\beta k^{2}}{2}\Delta \chi -
\frac{\alpha  h^{2}}{2}\Delta C =0.
                      \label{129}
\end{equation}

We solve (\ref{Emu}), (\ref{En}) and the $\delta_{\alpha\beta}$
part of (\ref{Eal}) as
\begin{equation}
                   \label{130}
\phi =\frac{\alpha h^{2}}{2}C  -\frac{\beta k^{2}}{2}\chi
\end{equation}
\begin{equation}
 A=uh^{2}C + vk^{2}\chi  ,
 \label{132}
\end{equation}
\begin{equation}
 F=-th^{2}C + vk^{2}\chi  ,
 \label{133}
\end{equation}
\begin{equation}
 B=-th^{2}C -wk^{2}\chi  ,
 \label{134}
\end{equation}

Substituting these expresions into (\ref{124})
we get  a relation on
 $\alpha$ and $\beta$
\begin{equation}
\alpha \beta=\frac{2qs}{q+r+s}
 \label{81m}
\end{equation}
and an expression for $h$
\begin{equation}
h=\pm\sqrt{\frac{4(q+r+s)}
{\alpha ^{2}(q+r+s)+2q(s+r)}},
 \label{819}
\end{equation}
Substituting these expresions into (\ref{125})
we get  the same relation  on
 $\alpha$ and $\beta$  as before as well as
\begin{equation}
k=\pm\frac{2\alpha (q+r+s)}{\sqrt{s
[2\alpha ^{2}(q+r)(q+r+s)+4q^{2}s]}}
 \label{195}
\end{equation}

By straitforward calculations one  can check that
the non-diagonal part of the Einstein equation is also satisfied under above conditions.

To summarize, the action (\ref{011}) has the solution of the form
(\ref{s2f}) expressed in terms
of two harmonic functions $H_{1}$ and $H_{2}$
if the parameters in the action are  related by (\ref{81m})
and the parameters in the Ansatz $h$ and $k$ are given
by (\ref{819}),(\ref{195}).


\subsection{Examples}
There is the relation (\ref{15}) between parameters $\alpha$
and $\beta$
in the action (\ref{011}). As a result the action
corresponds to the bosonic part of a supergravity theory
only in some dimensions.

If $D=4$ and $ q=d=1$ then
one can take $\alpha=\beta=1$ and the action
corresponds to the $SO(4)$ version of $N=4$ supergravity.
The solution  (\ref{12}) takes the form
\begin{equation}
 ds^{2}=-H_{1}^{-1}H_{2}^{-1}dy_{0}^{2}+H_{1}H_{2}dx^{\gamma}dx^{\gamma}
 \label{195a}
\end{equation}
This supersymmetric solution has been obtained in \cite{KLOPP}.

If $\alpha=\beta$ and $q=\tilde d$ then one has the solution
 \begin{equation}
 H_{2}^{-\frac{2}{q}}dz^{m}dz^{m}+
 dx^{\gamma}dx^{\gamma}],
  \label{713}
 \end{equation}
This solution was obtained in \cite{AV}. It contains
as a particular case for $d=10,~q=2$ the known solution
\cite{Ts212,CM}
$$  ds^{2}=H_{1}^{-\frac{3}{4}}H_{2}^{-\frac{1}{4}}
  (-dy_{0}^{2}+dy_{1}^{2})+$$
 $$ H_{1}^{\frac{1}{4}}H_{2}^{-\frac{1}{4}}
  (dz_{1}^{2}+dz_{2}^{2}+dz_{3}^{2}+dz_{4}^{2})+$$
\begin{equation}
  H_{1}^{\frac{1}{4}}H_{2}^{\frac{3}{4}}
  (dx_{1}^{2}+dx_{2}^{2}+dx_{3}^{2}+dx_{4}^{2}).
  \label{66m}
 \end{equation}
This solution has been used in the D-brane derivation
of the black hole entropy \cite{SV,CM}. Note however
that the solution (\ref{66m}) corresponds to the action
(\ref{011})  with the $3$-form $F_{3}$ and the $7$-form
$G_{7}$.


\section {Conclusion}
In conclusion, we have discribed multi-block p-brane solutions
for high dimensional gravity interacting with matter.
We have assumed the "electric" ansatz for the field and used
"no-force"
conditions for local fields (\ref{27}) together with the
harmonic gauge condition
(\ref{9}) to reduce  the system of differential equations
to a  system of non-linear algebraic equations.
The found solutions support a picture in which an
 extremal p-brane can be viewed as a composite of
`constituent' branes, each of the latter possessing a
corresponding charge.

\section*{Acknowledgments}

This work is supported by
an operating grant from the Natural Science and Engineering
Research Council of Canada. I.A.  and I.V. are
partially supported by the RFFI grants 96-01-00608 and
96-01-00312, respectively.

\end{document}